\documentclass[12pt,preprint]{aastex}

\DeclareGraphicsExtensions{.eps}

\begin{document}

\title{
High-energy $\gamma$-rays from  stellar associations }

\author{Diego F. Torres\altaffilmark{1},  Eva
Domingo-Santamar\'{\i}a\altaffilmark{2}, \& Gustavo E. Romero\altaffilmark{3}}

\altaffiltext{1}{Lawrence
Livermore National Laboratory, 7000 East Ave., L-413, Livermore,
CA 94550. E-mail: dtorres@igpp.ucllnl.org}
\altaffiltext{2}{Institut de F\'{\i}sica d'Altes
Energies (IFAE), Edifici C-n, Campus UAB, 08193 Bellaterra, Spain.
E-mail: domingo@ifae.es}
\altaffiltext{3}{Instituto
Argentino de Radioastronom\'{\i}a (IAR), C.C.\ 5, 1894 Villa
Elisa, Argentina. E-mail: romero@irma.iar.unlp.edu.ar}

\begin{abstract}
It is proposed that TeV $\gamma$-rays and neutrinos can be
produced by cosmic rays (CRs) through hadronic interactions in the
innermost parts of the winds of massive O and B stars. Convection
prevents low-energy particles from penetrating into the wind,
leading to an absence of MeV-GeV counterparts. It is argued that
groups of stars located close to the CR
acceleration sites in OB stellar associations may be detectable by
ground-based \v{C}erenkov telescopes.
\end{abstract}


\section{Introduction}

Several $\gamma$-ray sources are thought to be related
with early-type stars and their neighborhoods
(e.g., Montmerle 1979;
Cass\'{e} \& Paul 1980; Bykov \& Fleishman 1992a,b; Bykov 2001,
Romero \& Torres 2003). Recently, the first (and only)
TeV unidentified source was detected  in the Cygnus
region (Aharonian et al. 2002), where a nearby EGRET source (3EG
J2033+4118) has a likely stellar origin (White \& Chen 1992; Chen
et al. 1996; Romero et al. 1999; Benaglia et al. 2001). Here,
we explore whether CR illumination of
stellar winds of O and B stars can lead to
Galactic TeV $\gamma$-ray sources.

\section{The model}

O and B stars lose a significant fraction of their mass in stellar
winds with terminal velocities $V_\infty \sim 10^3$ km
s$^{-1}$. With mass loss rates as high as
$\dot{M}_\star=(10^{-6}-10^{-4})$ $M_{\sun}$ yr$^{-1}$, the
density at the base of the wind can reach $10^{-12}$ g cm$^{-3}$
(e.g., Lamers \& Cassinelli 1999, Ch.~2). Such winds are
permeated by significant magnetic fields, and provide a matter
field dense enough as to produce hadronic $\gamma$-ray emission
when pervaded by relativistic particles. A typical wind
configuration (Castor, McCray, \& Weaver 1975; V\"olk \&
Forman 1982; Lamers \& Cassinelli 1999, Ch.~12) contains an inner
region in free expansion (zone I) and a much larger hot compressed
wind (zone II). These are finally surrounded by a thin layer of
dense swept-up gas (zone III); the final interface with the
interstellar medium (ISM). The innermost region size is fixed by
requiring that at the end of the free expansion phase (about 100
years after the wind turns on) the swept-up material is comparable
to the mass in the driven wave from the wind, which happens at a
radius $R_{\rm wind}=V_\infty (3\dot{M}_\star/4 \pi \rho_0
V_\infty^3)^{1/2}$, where $\rho_0\approx m_p n_0$ is the ISM mass
density, with $m_p$ the mass of the proton and $n_0$ the ISM
number density. After hundreds of thousands of years, the wind
produces a bubble with a radius of the order of tens of parsecs,
with a density lower (except that in zone I) than in the ISM.
In what follows, we consider the hadronic production of
$\gamma$-rays in zone I, the innermost and densest region of the
wind. The matter there will be described through the
continuity equation: $\dot{M}_\star=4\pi r^2 \rho (r) V(r)$, where
$\rho(r)$ is the density of the wind and $V(r)=V_{\infty} (1-{R_0}/{r})^{\beta}$ is its velocity.
$V_{\infty}$ is the terminal wind velocity, and the parameter
$\beta$ is $\sim 1$ for massive stars (Lamers \& Cassinelli 1999,
Ch.~2). $R_0$ is given in terms of the wind velocity close to the
star, $V_0 \sim 10^{-2}V_\infty$, as $R_0=R_\star (1-(V_0/V_\infty)^{1/\beta})$.
Hence, the particle density
is $n(r)=\dot {M}_\star (1-{R_0}/{r} )^{-\beta}/({4\pi m_p
V_{\infty} r^2}). $


Not all CRs will enter into the
base of the wind. Although wind modulation has only been studied
in detail for the case of the relatively weak solar wind (e.g.
Parker 1958, Parker \& Jokipii 1970, K\'ota \& Jokipii 1983, Jokipii et al. 1993),
a first approach to determine whether particles can pervade the
wind is to compute the ratio ($\epsilon$) between the diffusion
and convection timescales: $t_d=3r^2/D$ and $t_c=3r/V(r)$,
where $D$ is the diffusion coefficient, and  $r$ is
the position in the wind. Only particles for which $\epsilon<1$
will be able to overcome convection and enter the dense wind
region to produce $\gamma$-rays through $pp$ interactions.
The diffusion coefficient is $D \sim \lambda_r c/3$, where
$\lambda_r$ is the mean-free-path for diffusion in the radial
direction.
As in White (1985) and
V\"olk and Forman (1982), the mean-free-path for scattering
parallel to the magnetic field ($B$) direction is assumed as $\lambda_\| \sim
10 r_g = 10 E/eB$, where $r_g$ is the particle gyro-radius and $E$
its energy. In the perpendicular direction $\lambda$ is shorter,
$\lambda_\bot \sim r_g$. The mean-free-path in the radial
direction is then given by $\lambda_r={\lambda_\bot}^2 \sin^2
\theta + {\lambda_\|}^2 \cos^2 \theta = r_g ( 10 \cos^2 \theta +
\sin^2 \theta)$, where $\cos^{-2} \theta = 1+(B_\phi/B_r)^2$.
Here, the geometry of the magnetic field is represented by the magnetic
rotator theory (Weber and Davis 1967; White 1985; Lamers and
Cassinelli 1999, Ch.~9) $B_\phi/B_r=(V_\star/V_\infty)
(1+r/R_\star)$ and $B_r=B_\star (R_\star/r)^2$, where $V_\star$ is
the rotational velocity at the surface of the star, and $B_\star$
the surface magnetic field.
Using all previous
formulae,
$\epsilon \sim \left\{ 3 e B_\star V_\infty (r-R_\star)(  {R_\star}/{ r})^2
( 1+( {V_\star }/{V_\infty} ( 1+  r/{R_\star} ) )^2
) ^{3/2} \right\} / \left\{ E_p c    (10+ ( { V_\star }/{V_\infty} ( 1+
 r/{R_\star} ) )^2 )  \right\}.$
%
%
The latter equation
defines a minimum energy $E_p^{\rm min}(r)$ below
which the particles are convected away from the wind (shown in
Fig. 1, left panel). Note that $E_p^{\rm min}(r)$ is an increasing
function of $r$, so particles that are not convected away in the
outer regions of the wind are able to diffuse up to its base.
$E_p^{\rm min}(r)$ can then be effectively approximated by
$E_p^{\rm min}(r\gg R_\star)$ in subsequent computations. Only
particles with energies higher than a few TeV will interact with
nuclei in the inner wind and ultimately generate $\gamma$-rays,
substantially reducing the flux in the MeV -- GeV band.

\begin{figure*}[t]
\centering
\includegraphics[width=0.75\textwidth,height=7cm]{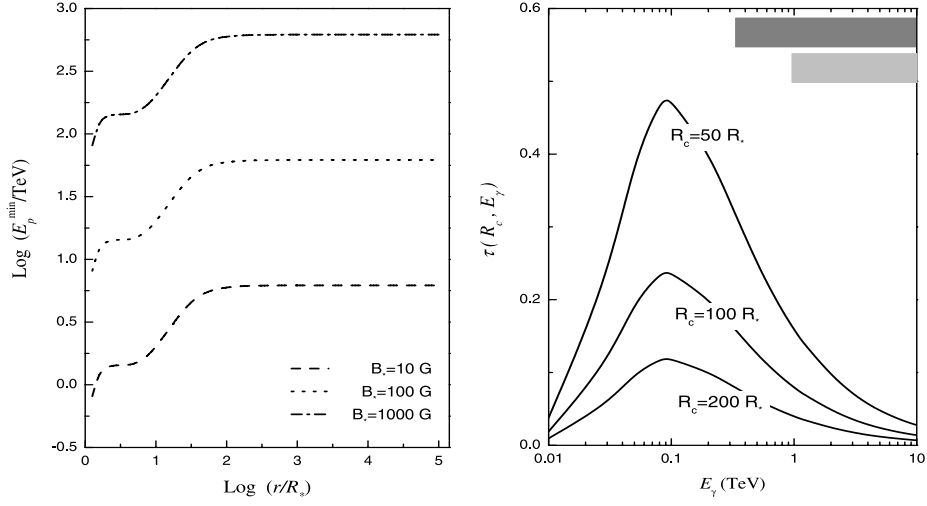}
\caption{Left: Minimum proton energy needed to overcome the wind
convection at different distances from the star. Here
$V_{\star}=250$ km s$^{-1}$, $V_{\infty}=1750$ km s$^{-1}$,
$R_\star=12\,R_\odot$. Right: Opacities to pair production as a
function of the $\gamma$-ray energy for different creation
places $R_c$. Here, the star has $T_{\rm eff}=38000$ K.
}
\end{figure*}


The opacity to pair production of the $\gamma$-rays in the UV
stellar photon field can be computed as
$\tau(R_{c},E_\gamma)=\int_0^\infty \int_{R_c}^\infty N(E_\star)
\sigma_{e^-e^+}(E_\star,E_\gamma) dE_\star dr$, where $E_\star$ is
the energy of the photons emitted by the star, $E_\gamma$ is the
energy of the $\gamma$-ray, $R_c$ is the place where the photon
was created within the wind, and
$\sigma_{e^-e^+}(E_\star,E_\gamma)$
is the cross section for $\gamma\gamma$ pair production (Cox 1999,
p.214). The stellar photon distribution is that of a blackbody
peaking at typical star effective temperatures ($T_{\rm eff}$),
$N(E_\star)=(\pi B(E_\star)/h E_\star c) R_\star^2/r^2$, where $h$
is the Planck constant, and
$B(E_\star)=(2{E_\star}^3/(hc)^2)/(e^{E_\star/kT_{\rm eff}} -1)$.
$\tau(R_{c},E_\gamma)$ is shown in Fig. 1 (right panel) for
different photon creation sites ($R_c \ll R_{\rm wind}$).
$\gamma$-ray photons of TeV and higher energies do not encounter
significant opacities in their way out of the wind, unless they
are created at its very base, hovering over the star (which is
unlikely to happen because $R_{\rm wind} \gg R_\star$ and the
proton propagates in a high magnetic field environment).
Although we show the opacity for values of the photon energy as
low as 100 GeV, most of the $\gamma$-rays will have higher
energies, since only protons with $E_p
> E_p^{\rm min}$ will enter the wind. The grey (light-grey) box
in the figure shows typical energies of $\gamma$-rays for the case of
a surface magnetic field $B_\star=10$ G (100 G). There is a
large uncertainty about the typical values for the
magnetic field in the surface of O and B stars, but recent
measurements favor $B_\star \gtrsim 100$ G (e.g., Donati et al.
2001; 2002).




\section{$\gamma$-ray and neutrino emission}

The differential $\gamma$-ray emissivity from $\pi^0$-decays can
be approximated by $q_{\gamma}(E_{\gamma})= 4 \pi \sigma_{pp}(E_p)
({2Z^{(\alpha)}_{p\rightarrow\pi^0}}/{\alpha})\;J_p(E_{\gamma})
\eta_{\rm A}\Theta(E_p-E_p^{\rm min})$ at the energies of
interest.  The parameter $\eta_{\rm A}$ takes into account the
contribution from different nuclei in the wind (for a standard
composition $\eta_{\rm A} \sim 1.5$, Dermer 1986).
$J_p(E_{\gamma})$ is the proton flux distribution evaluated at
$E=E_{\gamma}$ (units of protons per unit time, solid angle,
energy-band, and area). The cross section $\sigma_{pp}(E_p)$ for
$pp$ interactions at energy $E_p\approx 10 E_{\gamma}$ can be
represented above $E_p\approx 10$ GeV by $\sigma_{pp}(E_p)\approx
30 \times [0.95 + 0.06 \log (E_p/{\rm GeV})]$ mb (e.g., Aharonian
\& Atoyan 1996). $Z^{(\alpha)}_{p\rightarrow\pi^0}$ is the
so-called spectrum-weighted moment of the inclusive cross-section.
Its value for different spectral indices $\alpha$ is given, for
instance, by
Drury et al. (1994). Finally $\Theta(E_p-E_p^{\rm min})$ is a
Heaviside function that takes into account the fact that only CRs
with energies higher than $E_p^{\rm min}(r\gg R_\star)$ will
diffuse into the wind.  The spectral $\gamma$-ray intensity
(photons per unit time and energy-band) is $
I_{\gamma}(E_{\gamma})=\int n(r) q_{\gamma}(E_\gamma) dV, $ where
$V$ is the interaction volume. The luminosity in a given band is
$L_\gamma=\int_{R_\star}^{R_{\rm wind }} \int_{E_1}^{E_2} n(r)\,
q_{\gamma}(E_\gamma) E_\gamma \, (4\pi r^2)dr \, dE_\gamma $
(e.g. Torres et al. 2003, Romero et al. 2003 for details). Assuming
a canonical spectrum for the relativistic CR population,
$J_p(E_p)=(c/4\pi) N(E_p)=(c/4\pi) K_p {E_p}^{-\alpha}$, the result
(in the range $E_\gamma \sim 1-20$ TeV)
can be expressed in terms of the normalization $K_p$ and will
depend on all other model parameters, mainly on the proton
(photon) spectral index, the ISM density, the
terminal velocity, and the mass-loss rate. Very mild dependencies appear with $\beta$ and
$R_\star$. Table 1 presents results for the luminosity
for typical values of all these parameters. We have fixed
$\dot {M}_\star = 10^{-5} M_\odot$ yr$^{-1}$, $\beta=1$, and
$R_\star=12 R_\odot$ in this example. The mass contained in the
innermost region of the wind, $M_{\rm wind}$, is also shown.
$L_\gamma \sim 10^{25-30} K_p$ erg
s$^{-1}$ can be obtained as the luminosity produced by one
particular star; the total luminosity of a group of stars should
add contributions from all illuminated winds. Convolving the
previous integral with the probability of escape (obtained through
the opacity as $e^{-\tau}$) does not noticeably change these results.
Finally, it is possible to factor out the
normalization in favor of the CR enhancement in the region where
the wind is immersed. The CR energy density is $ \omega_{\rm CR} =
\int N_p(E_p) E_p dE_p = 9.9 K_p\; {\rm eV}\; {\rm cm}^{-3} \equiv
\varsigma \omega_{\rm CR,\odot} , $ where $\varsigma$ is the
enhancement factor of the CR energy density with respect to the
local value, $ \omega_{\rm CR,\odot}$
(energies between 1 GeV and 20 TeV). Then, $K_p \sim (0.2-0.3)
\varsigma $.

\begin{table*}[t]
\caption{Examples for hadronic $\gamma$-ray luminosities from
typical stellar wind configurations. }
\begin{center}
\begin{tabular}{ccrrrccc} \hline

Model&  $V_\infty$ & $n_0$     &  $R_{\rm wind}$ & $M_{\rm wind}$       & $L^{\alpha=1.9}_{\gamma}/ K_p $ & $L^{\alpha=2.0}_{\gamma} / K_p $ & $L^{\alpha=2.1}_{\gamma}/ K_p $ \\ 
     & (km s$^{-1}$) & (cm$^{-3}$) &   (pc)            &  ($M_\odot$) &   (erg s$^{-1}$)  & (erg s$^{-1}$)    &  (erg s$^{-1}$) \\ 
\hline \hline
a & 1750 & 10   & 0.07 & 0.0004 & 2 $\times 10^{28}$ & 7 $\times 10^{26}$ & 3 $\times 10^{25}$   \\ 
b & $\cdots$  & 1    & 0.24 & 0.0013  & 5 $\times 10^{28}$ & 2 $\times 10^{27}$ & 8 $\times 10^{25}$  \\ 
c & $\cdots$  & 0.1  & 0.75 & 0.0041  & 2 $\times 10^{29}$ & 7 $\times 10^{27}$ & 3 $\times 10^{26}$  \\ 
d & $\cdots$  & 0.01 & 2.4  & 0.0130  & 5 $\times 10^{29}$ & 2 $\times 10^{28}$ & 8 $\times 10^{26}$  \\ 
e & 1000 & 10   & 0.09 & 0.0009 & 4 $\times 10^{28}$ & 1 $\times 10^{27}$ & 6 $\times 10^{25}$   \\ 
f & $\cdots$  & 1    & 0.31 & 0.0030  & 1 $\times 10^{29}$ & 5 $\times 10^{27}$ & 2 $\times 10^{26}$  \\ 
g & $\cdots$  & 0.1  & 0.99 & 0.0095   & 4 $\times 10^{29}$ & 1 $\times 10^{28}$ & 6 $\times 10^{26}$ \\ 
h & $\cdots$  & 0.01 & 3.1  & 0.0301   & 1 $\times 10^{30}$ & 5 $\times 10^{28}$ & 2 $\times 10^{27}$ \\ 
i & 800  & 10   & 0.11 & 0.0013  & 5 $\times 10^{28}$ & 2 $\times 10^{27}$ & 9 $\times 10^{25}$  \\ 
j & $\cdots$  & 1    & 0.35 & 0.0042   & 1 $\times 10^{29}$ & 7 $\times 10^{27}$ & 3 $\times 10^{26}$ \\ 
k & $\cdots$  & 0.1  & 1.1  & 0.0133    & 5 $\times 10^{29}$ & 2 $\times 10^{28}$ & 9 $\times 10^{26}$\\ 
l & $\cdots$  & 0.01 & 3.5  & 0.0421    & 1 $\times 10^{30}$ & 7 $\times 10^{28}$ & 3 $\times 10^{27}$\\ 
\hline
\end{tabular}
\end{center}
\end{table*}


The $\nu_\mu + \bar\nu_\mu$ neutrino flux ($F_\nu(E_\nu )$) will
be derived from the observed $\gamma$-ray flux
($F_\gamma(E_\gamma)$) by imposing energy conservation
(see Alvarez-Mu\~niz \& Halzen 2002 for details):
$ \int
E_\gamma F_\gamma(E_\gamma) dE_\gamma = C \int
E_\nu F_\nu(E_\nu ) dE_\nu, $ where the limits of the integrals
are ${E_{\gamma\; [\nu]}^{\rm min}}$ ($E_{\gamma \; [\nu]}^{\rm
max}$), the minimum (maximum) energy of the photons [neutrinos]
and the pre-factor $C$ is a numerical constant of order one.
Using the resulting $\nu$  flux, the signal for the detection of
$\nu$-events  can
be approximated  as (Anchordoqui et al. 2003) $S = T_{\rm
obs}  \int dE_\nu A_{\rm eff} F_\nu(E_\nu) P_{\nu \to \mu} (E_\nu)
$ whereas the noise will be given by $ N={\left(
 T_{\rm obs}
  \int dE_\nu  A_{\rm eff} F_{\rm B}(E_\nu) P_{\nu \to \mu} (E_\nu)
 \Delta \Omega
\right)^{(1/2)}}, $ where$\Delta \Omega$ is the solid angle of the
search bin ($\Delta \Omega_{1^\circ \times 1^\circ} \approx 3
\times 10^{-4}$~sr for ICECUBE,  Karle 2002)
 and
$F_{\rm B}(E_\nu) \lesssim 0.2 \,(E_\nu/{\rm
GeV})^{-3.21}$~GeV$^{-1}$ cm$^{-2}$ s$^{-1}$ sr$^{-1}$  is the
$\nu_\mu + \bar \nu_\mu$ atmospheric $\nu$-flux~(Volkova 1980,
Lipari 1993). Here, $P_{\nu \to \mu} (E_\nu) \approx 3.3 \times
10^{-13} \,(E_\nu/{\rm GeV)}^{2.2}$ denotes the probability that a
$\nu$ of energy $E_\nu \sim 1 - 10^3~{\rm GeV}$, on a trajectory
through the detector, produces a muon~(Gaisser et al. 1995).
$T_{\rm obs}$ is the observing time and $A_{\rm eff}$ the
effective area of the detector. Those systems producing a
detectable $\gamma$-ray flux above 1 TeV are prime candidates to
also be detectable neutrino sources, see below.

\section{Source location and luminosity}


The flux expected at Earth from an isolated star can be computed
as $F_\gamma (E_\gamma>1 {\rm TeV}) =(1/ 4\pi D^2) \int_{R_\star}^{R_{\rm wind }}
\int_{1 {\rm TeV}} n(r)\, q_{\gamma}(E_\gamma) \, 4\pi r^2 \, dr
\, dE_\gamma.$ The models in Table 1, at 2 kpc,  give fluxes in
the range $(1\times10^{-20} - 7 \times 10^{-16}) K_p$ photons
cm$^{-2}$ s$^{-1}$. Hence, there are models for which a small
group of $\sim$ 10 stars in a region with a CR enhancement factor
of $\sim 100$ might be detectable at the level of $\sim 10^{-14}$
photons cm$^{-2}$ s$^{-1}$.

CRs are expected to be accelerated in OB associations through
turbulent motions and collective effects of stellar winds (e.g.
Bykov \& Fleishman 1992,b). The main acceleration region for TeV
particles would be in the outer boundary of the supperbubble
produced by the core of the association. If there is a subgroup of
stars located at the acceleration region, their winds might be
illuminated by the locally accelerated protons, which would have a
distribution with a slope close to the canonical value,
$\alpha\sim2$. For stars out of the acceleration region, the
changes introduced in the proton distribution by the diffusion of
the particles (a steepening of its spectrum) would render the
mechanism for TeV $\gamma$-ray production inefficient. This can be
seen from Table 1 through the strong dependency of the predicted
TeV luminosity on the spectral slope of the particles.


An important assumption in our model is that the diffusion
coefficient is a linear function of the particle energy in the
inner wind. This is indeed an assumption also in both V\"olk
\& Forman (1982) and White's (1985) models of the particle
diffusion in the strong winds of early-type stars, among other
studies. Measurements of the solar wind, however, seem to suggest
a harder relation with energy (e.g., $D\propto E^{0.4-0.5}$,
Ginzburg \& Syrovatskii 1964, p.336). If such a relation would
hold for the inner wind of an O star (where $pp$ interactions
occur), depending on the constant of proportionality, it
could yield a higher value of $E_{p}^{\rm min}$ and
hence a lower $\gamma$-ray luminosity. However, contrary to what
happens with the Sun, in early-type stars line-driven
instabilities are expected to produce strong shocks in the inner
wind (Lamers \& Cassinelli 1999). In such a scenario, as
emphasized by White (1985), to expect that particles will diffuse
according to the Bohm parameterization seems not to be
unreasonable. As we discuss in the next section, direct
observation of TeV sources of stellar origin can shed light on the
issue.



\section{Application: the unidentified TeV source }

The  HEGRA detection in the vicinity of Cygnus OB2, TeV J2032+4131
(Aharonian et al. 2002), presents an integral flux
$F_\gamma(E_\gamma>1$TeV$) = 4.5(\pm 1.3) \times
10^{-13}$ photons cm$^{-2}$ s$^{-1}$, and
a $\gamma$-ray spectrum  $ F_\gamma(E_\gamma)= B ({ E_\gamma}
  /{{\rm TeV}})^{-\Gamma} $  photons cm$^{-2}$ s$^{-1}$ TeV$^{-1}$,
 where  $ B       =  \,4.7\, (\pm2.1_{\rm stat}
 \pm1.3_{\rm sys}) \times 10^{-13} $ and $
 \Gamma  = 1.9  (\pm0.3_{\rm stat} \pm0.3_{\rm sys}).  $
No counterparts at lower energies are presently identified (Butt
et al. 2003, Mukherjee et al. 2003). The source flux was constant
during the three years of data collection. The extension of the
source (5.6$\pm 1.7$ arcmin)
disfavors a pulsar or active galactic nuclei origin. The absence of an X-ray
counterpart additionally disfavors a microquasar origin. Instead,
the location of the TeV source, separate from the core of the
association, and coincident with a significant enhancement of the
star number density (see Fig. 1 of Butt et al. 2003) might suggest
the scenario outlined in the previous section.

A nearby EGRET source is, on the other hand, coincident with the
center of the association, where it might be produced either in
the terminal shocks of powerful stars therein existing (White and
Chen 1992, Chen et al. 1996), or in the colliding wind binary
system Cyg OB2 \#5 (Benaglia et al. 2001), or in a combination of
these scenarios. Contributions from the inner winds of OB
stars as in the model herein explored cannot be ruled out.
These, however, are not expected to dominate because of wind modulation (at low energies) and
of the softening of the CR spectrum while diffusing
from the superbubble accelerating region, which
significantly diminish the number of $pp$ interactions in the
winds.

Our model could explain the unidentified TeV source without
requirements other than the presence of the already observed stars
and a reasonable CR enhancement if the density of the original ISM
was rather low. Butt et al. (2003) argued for a density of
$n_0=30$ cm$^{-3}$. However,  this
should be taken as a generous upper limit:  a) Apparently, there is no star formation
currently active at the position of the source. b) The CO-H$_2$
conversion factor used to compute the density has been taken as
the normal Galactic one, but it could be lower
in the neighborhood of star forming environments (e.g., Yao
et al. 2003). c) The particle density within the TeV source region
has been averaged from a velocity range integrated along the line
of sight corresponding to 3700 pc and including the core of the
Cygnus association. d) The TeV source will actually be immersed in
the zone II of the winds of the several powerful stars therein
detected, which should have swept the ISM away and diminished its
density. Our models (e.g., model g of Table 1), which in fact
take for the stellar parameters an average value from the stars in
Table 3 of Butt et al. (2003), show that the illumination of the
innermost regions of the winds of $\sim 10$ stars with a CR
enhancement of $\sim 300$ in a medium density of about 0.1
cm$^{-3}$ may be enough to produce the HEGRA detection.
The neutrino flux that results from a hadronic
production of the TeV $\gamma$-ray source would not produce a
significant detection in AMANDA II, which is consistent with
the latest reports by the AMANDA collaboration (Ahrens et al.
2003).
In ICECUBE, however, the signal-to-noise is $\sim 1.8$ for 1~yr of
observation (for energies above 1 TeV, an effective area of 1
km$^2$, before taking into account neutrino oscillations effects).
If ICECUBE can reach a $1^\circ \times 1^\circ$ or finer search
bin, and a km$^2$ effective area at TeV energies, a long
integrating time could distinguish the hadronic origin of the
HEGRA detection.

\section{Concluding Remarks}

Hadronic interactions within the innermost region of the winds of
O and B stars can produce significant $\gamma$-ray luminosities at
TeV energies, with low brightness at other energies. At distances
less than a few kpc, several illuminated winds pertaining to
subgroups of stars located at CR acceleration regions in OB
associations might be detected by
\v{C}erenkov telescopes.
A reasonable set of model parameters can be found to produce a
flux compatible with the only unidentified TeV source known.
A candidate selection for possible new TeV sources, based on
these predictions, will be reported elsewhere.

We thank L. Anchordoqui, P. Benaglia, Y. Butt, C. Mauche, F.
Miniatti, R. Porrata, and H. V\"olk for useful discussions. The
work of DFT was performed under the auspices of the US DOE (NNSA),
by UC's LLNL under contract No. W-7405-Eng-48. ED-S acknowledges
the Ministry of Science and Technology of Spain for financial
support and the IGPP/LLNL for hospitality. GER is mainly supported
by Fundaci\'on Antorchas, and additionally, from grants PICT
03-04881 (ANPCyT) and PIP 0438/98 (CONICET). He is grateful to the
Hong Kong University and Prof. K.S. Cheng for hospitality.


\end{document}